\documentclass[aps,reprint,showpacs,superscriptaddress,
footinbib,citeautoscript]{revtex4-1}

\usepackage{graphicx}

\usepackage[utf8]{inputenc}
\usepackage{csquotes}
\usepackage[american]{babel}
\usepackage[T1]{fontenc}
\usepackage{enumerate}
\usepackage{mdwlist}
\usepackage{color}
\usepackage{amssymb}
\usepackage{amsmath}
\usepackage{amsfonts}
\usepackage{mathrsfs}

\usepackage{bm}
\usepackage{dcolumn}
\usepackage{color}
\usepackage[colorlinks=true,citecolor=blue]{hyperref}
\hypersetup{colorlinks=true,citecolor=blue,linkcolor=red,urlcolor=blue}

\usepackage{amsmath,amssymb}
\usepackage{color}
\usepackage{graphicx}
\usepackage{natbib}

\begin{document}
\title{Nonadiabatic transition at a band-touching point}
\author{Mohammad-Sadegh Vaezi}
\affiliation{Passargad Institute for Adavanced  Innovative Solutions (PIAIS), Tehran 19395-5531, Iran}
\affiliation{Department of Converging Technologies, Khatam University, Tehran 19395-5531, Iran}
\author{Davoud Nasr Esfahani}
\email{d.nasr@khatam.ac.ir}
\email{dd.nasr@gmail.com}
\affiliation{Department of Converging Technologies, Khatam University, Tehran 19395-5531, Iran}
\affiliation{Passargad Institute for Adavanced  Innovative Solutions (PIAIS), Tehran 19395-5531, Iran}

\date{\today}

 %%%%%%%%%%%%%%%%%%%%%%%%%%%%%%%%%%%%%%%%%%%%%%%%%%%%%%%
 %=========================================================================
 %%%%%%%%%%%%%%%%%%%%%%%%%%%%%%%%%%%%%%%%%%%%%%%%%%%%%%%

\begin{abstract}
Low-energy Hamiltonians with a linear crossing in their energy dispersion (dubbed Dirac Hamiltonians) have recently been the subject of intense investigations. 
 The linear dispersion is often the result of an approximation in the energy dispersion at the band-touching point in which higher order terms are discarded. 
 In this paper, we show that, in terms of nonadiabatic transitions, by passing through a touching point, certain types of quadratic terms could not be omitted, even in the arbitrary vicinity of it, i.e., quadratic terms could significantly affect the transition probability, hence the Hamiltonian is not reducible to a linear one. We further show that the presence of terms with exponents larger than two only affects the transition probability away from the touching point. In
the end, we discuss conditions that may lead to the appearance of oscillations in the transition probability profile.
\end{abstract}

\maketitle

 %%%%%%%%%%%%%%%%%%%%%%%%%%%%%%%%%%%%%%%%%%%%%%%%%%%%%%%
 %=========================================================================
 %%%%%%%%%%%%%%%%%%%%%%%%%%%%%%%%%%%%%%%%%%%%%%%%%%%%%%%

\section{Introduction}
  \indent Two-level systems (TLSs) are among the simplest quantum models due to their two-dimensional Hilbertexcited space and manifest
 in a variety of physical phenomena \cite{doi:10.1142/4783,SHEVCHENKO20101,Heyl_2018,Graafeabc5055}, some of which intrinsically have only two levels (e.g., spin half systems).
 Recently, they have gained considerable attention due to their applicability for simulation of quantum bits in quantum simulators and feasible applications in future quantum computation devices\cite{PhysRevA.74.052330,RevModPhys.86.153,PhysRevLett.97.077001,M_ller_2019}.
 One of the well-known TLSs that has been the subject of considerable investigation
 is the Landau-Zener (LZ) model, which describes a gapped system such that the crossing of the simple linear energy dispersion is avoided \cite{landau1932,stuckelberg1932,zener1937,PhysRevA.74.052330}.
 %Originally, LZ Hamiltonian was introduced to tackle the problem of atomic collisions which apparently has an avoided crossing (AC) character.
%Since the invention of the LZ Hamiltonian, it has been the subject of considerable works. 
In this model, the transition probability (TP) between the ground-state and the excited state 
 ${\cal P}_{LZ}= e^{-\pi\Delta^2/c}$, where $\Delta$ is  the energy gap, and $c$ is the speed of the driving of the system. \\ 
 \indent Within the context of the nonadiabatic transitions, so far, most of the attention associated with TLSs has been focused on gapped systems\cite{SHEVCHENKO20101,PhysRevLett.110.016603,PhysRevA.90.062120,PhysRevB.96.054301,PhysRevB.97.035428,PhysRevB.103.144301,PhysRevA.86.033415,PhysRevA.59.4580,PhysRevA.88.043404,suominen1992parabolic,PhysRevA.86.063613,PhysRevB.99.205426}. 
 On the other hand, there are a variety of physically relevant systems which are gap-less through the presence of a touching point in the energy dispersion, e.g., graphene\cite{novoselov2005,geim2007,castro2009,Goerbig2017}, 
  1D and 2D p-wave superconductors
  \cite{Kitaev2001,volovik1999,bernevig2013topological}, and semi-Dirac materials\cite{pardo2009,huang2015,banerjee2012,mili2019}. 
  Recently, tun-able construction of such gap-less points has been realized in optical lattices\cite{tarruell2012,polini2013,uehlinger2013} which are reasonable platforms for quantum simulators \cite{PRXQuantum.2.017003}. 
   Noticeably, for the Dirac points the low-energy Hamiltonian could be well approximated with TLSs by considering only the linear terms\cite{Montambaux2009,katsnelson_2012}.
   Nonadiabatic processes across such kind of cones have been studied in 
    the optical lattice experiments\cite{lim2012,sun2018,lim2014,xu2014,uehlinger2013_2}, where away from the touching points any intersection of a plane parallel to the energy and y axes appears to be an avoided crossing (AC)\cite{lim2012}. 
    Therefore, any transition from the ground-state to the excited state could be  explained based on the LZ formula in this regime (i.e., away from touching points).\\
  \indent For a linear crossing problem, where there is no mixing between diabatic levels (which is a gap-less problem), the transition to the excited state ${\cal P}_{LZ}=1$. Similarly, by nonadiabatic passage through the Dirac points, the transition to the excited state should be equal to one \cite{lim2012}.  \\
  \indent We know that in realistic physical systems a perfect linear crossing does not exist. 
  That is, linear models are low-energy Hamiltonians where higher order terms are neglected at the proximity of coalescence points. 
  In this paper, we show that in the context of nonadiabatic transitions, such Hamiltonians are not reducible to a linear one (i.e., no matter how close we are to the the touching point in the parameter space). To this end, we employ the concept of fidelity susceptibility (FIS), hereafter denoted by $\chi$\cite{you2007}. We consider a Hamiltonian $\hat{H}(\lambda)$ with a touching point at $\lambda=0$. By expanding $\hat{H}(\lambda)$ around $\lambda=0$, we prove that if certain quadratic terms are neglected, $\chi(\lambda)\rightarrow 0$ as $\lambda\rightarrow 0$, while keeping them results in a finite $\chi(\lambda)$ as $\lambda\rightarrow 0$. We then show that such non-vanishing $\chi(0)$ leads to 
   a TP less than one upon nonadiabatic passage through the touching point. Furthermore, we discuss the TP far away from the touching point based on FIS.\\
    \indent The outline of this paper is as follows. 
    In Sec.~\ref{method} we briefly review the concept of FIS and its relation with the TP. 
    In Sec.~\ref{models}, we investigate FIS for two interesting gapless models followed by relevant discussions. Qualitative analyses and numerical results are provided  in Secs.~\ref{analysis} and \ref{numtest}, respectively. We summarize our main message and findings in Sec.\ref{conclusion}.
       
 %%%%%%%%%%%%%%%%%%%%%%%%%%%%%%%%%%%%%%%%%%%%%%%%%%%%%%%
 %=========================================================================
 %%%%%%%%%%%%%%%%%%%%%%%%%%%%%%%%%%%%%%%%%%%%%%%%%%%%%%%
      
\section{Fidelity susceptibility and nonadiabatic transitions}\label{method}
Let us consider a TLS which is explained based on a Hamiltonian $\hat{H}(\lambda)$, where $\lambda$ is a real scalar. 
 A convenient way for investigation of the time evolution of such systems, with a discreet basis set, is to represent the evolved wave function based on the instantaneous basis set, i.e., $|\psi(t)\rangle=\sum_{n=0,1} c_n(t)|n(\lambda(t))\rangle$, 
where $|n(\lambda(t)\rangle$s are the eigenstates of the following time-independent Schrodinger equation $H(\lambda(t))|n_\lambda(t)\rangle=E_n(t)|n_\lambda(t)\rangle$, and $t$ denotes time. 
 In this representation the equation of motion for each $c_n(t)$ reads\cite{tomka2012}, 
 \begin{equation}\label{c2-eq21}
   \dot{\tilde{c}}_n(t) = -\sum_{m\neq n} e^{i\theta_{nm}(t)} \tilde{c}_m(t) \dot{\lambda}\frac{\langle
n_\lambda(t)|\partial_\lambda{H(\lambda(t))}|m_\lambda(t)\rangle}{(E_m-E_n)},
 \end{equation}
 where $\theta_{nm}(t)= \theta_n(t)-\theta_m(t)$, $\theta_n(t) = \int_0^tE_n(\tau)d\tau - i\int_0^t \langle n_\lambda(\tau)|\dot{n}_\lambda(\tau)\rangle d\tau$, and $\tilde{c}_n(t)=c_n(t)e^{i\theta_{n}(t)}$. 
 Therefore, starting from one of the eigenstates of the Hamiltonian, if the system is driven by changing $\lambda$ with a pace of $\dot{\lambda}$, one may expect that larger 
   $\dot{\lambda}\frac{\langle n_\lambda(t)|\partial_\lambda{H(\lambda (t))}|m_\lambda(t)\rangle}{(E_m-E_n)}$ would acquire larger transition to the other state.\\

\indent For TLSs the absolute value of $\frac{\langle n_\lambda(t)|\partial_\lambda{H(\lambda (t))}|m_\lambda(t)\rangle}{(E_m-E_n)}$  with $n\neq m$ is nothing but the square root of FIS associated with the $n$th level, i.e., $\chi_n(\lambda)$. Here $\chi_0(\lambda)=\chi_1(\lambda)=\chi(\lambda)$ holds. \\
\indent It is known that FIS can be employed as a general measure of phase transitions \cite{you2007,troyer2}. 
Moreover, FIS has been utilized as a qualitative upper bond for TP \cite{PhysRevE.79.061125}, and is instrumental for the comparison of ground-state probability of driven systems with different ground-state orders.\cite{PhysRevB.90.205121} \\
\indent When there is a gap in the system, semi-classical analysis is a well-established method for the estimation of a TP. In particular, the Dykhne-Davis-Pechukas (DDP)\cite{dykhne1960quantum,davis1976nonadiabatic} method  widely has been employed for the estimation of nonadiabatic TPs of TLSs\cite{suominen1992parabolic,PhysRevA.59.4580,PhysRevA.86.033415}. While DDP seems to be very successful, it has two drawbacks. First,  it works only for gapped models, i.e., the instantaneous eigenstates of the model must not cross in real time\cite{dykhne1960quantum,davis1976nonadiabatic,suominen1992parabolic,PhysRevA.59.4580,PhysRevA.86.033415} (yet they may cross in complex time domain). Second, it gives an estimate of TP for a driving $t=-\infty \to  +\infty$. Hence, one may lose information about intermediate transitions throughout the evolution. Fortunately, by employing often low-cost numerical integration of time-dependent Schrodinger equation for TLSs, one may achieve rather accurate results for the magnitude of TPs for both gapped and gap-less models. \\
\indent Based on the above-mentioned discussions, since we mainly focus on gap-less models, in the rest of this paper we will use FIS to detect active regions of the phase space where a transition to the other state is more likely. We should note that due to the role of $\theta_{mn}$ (which is related to the band structure and geometrical phase \cite{tomka2012})  in Eq.~(\ref{c2-eq21}), if two systems  have the same FIS, we cannot necessarily assert that they have the same nonadiabatic behavior.\\
  \indent Throughout the rest of this paper, we will suppose that all of the systems are driven
   such that  $\lambda(t)=c t $.  To numerically evolve $|\psi(t)\rangle$ in time, we will employ the Crank-Nikelson method \cite{oka2003}. 
   For all numerical simulations we consider $\hbar=1$ and set the time step equal to 0.001.\\
   \indent By starting from some $|\phi_0(\lambda(t_i))\rangle$, we define the TP $\displaystyle {\cal P}=|\langle \phi_{1}(\lambda(t_f)) |\psi(t_f)\rangle|^2$ at specific final $t_f$. The $\lambda(t_i)$ and $\lambda(t_f)$ are fixed by the protocol of driving. The $0$ and $1$ subscripts denote the instantaneous ground-state and excited state, respectively. \\
   \indent We further define the following protocols of driving of the Hamiltonians under study: PL$_1$ corresponds to a driving $-\lambda_0\rightarrow \lambda_0$, where $\lambda_0$ is a starting point which is very close to touching point. Here we fix $\lambda_0=0.1$. PL$_{-\infty,0}$ corresponds to a driving $\lambda(-\infty) \rightarrow \lambda(0)$ and PL$_{0,\infty}$ corresponds to a driving from $\lambda(0)\rightarrow \lambda(+\infty)$. PL$_2$ is assigned for a driving $\lambda(-\infty) \rightarrow \lambda(+\infty)$. \\
    % $t=+\infty$, where $|\phi_{+}(t)\rangle$ and $|\psi(t)\rangle$ are the instantaneous excited state and the evolved state, respectively.
   
%%%%%%%%%%%%%%%%%%%%%%%%%%%%%%%%%%%%%%%%%%%%%%%%%%%%%%%
 %=========================================================================
 %%%%%%%%%%%%%%%%%%%%%%%%%%%%%%%%%%%%%%%%%%%%%%%%%%%%%%%
   
\section{Models and Discussions}\label{models}
  We start with a gap-less Hamiltonian corresponding to the boundary of a trivial and topological phase of a p-wave superconductor \cite{bernevig2013topological},
  \begin{eqnarray}
{ H}_{{PW}}(\lambda)= \displaystyle\left( {\begin{array}{cc} \frac{\lambda^2}{2m} & \lambda\Delta\\ \lambda\Delta^* &  -\frac{\lambda^2}{2m}\\ \end{array} } \right),\nonumber
\end{eqnarray}
  where $\lambda$ is momentum, $m$ mass of electron, and $\lambda\Delta$ is the linear coupling. The  energy  dispersion $\displaystyle E(\lambda)=\pm|\lambda|\sqrt{\frac{\lambda^2}{4m^2}+|\Delta|^2}$,  with a touching point at $\lambda=0$. The corresponding FIS reads,
  \begin{equation}
   \chi(\lambda) =  \left ( \frac{2m\Delta}{(2m\Delta)^2+\lambda^2} \right )^2,
  \end{equation}
with the property, $\lim_{\lambda\rightarrow 0}\chi(\lambda)= 1/4m^2\Delta^2$. For finite values of $m$, one may be tempted to ignore the diagonal terms in the Hamiltonian for small enough values of $\lambda$ near $0$, which results in an approximated Hamiltonian $\tilde{H}_{PW}(\lambda)= \displaystyle\left( {\begin{array}{cc} 0 & \lambda\Delta  \\  \lambda\Delta^* & 0 \\ \end{array} } \right)$ with $\tilde{\chi}(\lambda) = 0$. Here, the two diabatic bands become completely unperturbed. 
 However, this is in stark difference with what we found earlier for FIS, i.e.,  $\lambda\rightarrow 0$, $\chi(\lambda)\rightarrow 1/4m^2\Delta^2$. Based on our discussions in Sec.\ref{method}, this shows that, in terms of time evolution of the system, $H_{PW}$ is not reducible to a simple linear model even for very small 
  values of $\lambda$. It is worth mentioning that only when $m \to \infty$, ignoring the quadratic terms is a valid approximation.\\
\indent As an another instructive example, we consider the low-energy Hamiltonian for a graphene-like system with a prominent property that acquires a gap-less point at $\boldsymbol{K}=\frac{2 \pi}{3 a}\left(1, \frac{1}{\sqrt{3}}\right)$. 
 The tight-binding Hamiltonian is given by \cite{katsnelson_2012}
 \begin{eqnarray}
 H_{GR}=\displaystyle\left( {\begin{array}{cc} 0 & s(\boldsymbol{k}) \\  s^*(\boldsymbol{k}) & 0 \\ \end{array} } \right),
 \end{eqnarray}
 where $s(\boldsymbol{k})=-h e^{-i k_{x} a}\left(1+2 e^{\left(i \frac{3 k_{x} a}{2}\right)} \cos \frac{\sqrt{3}}{2} k_{y} a\right)$, $h$ is the hopping parameter, and $a$ is the lattice constant.
 Often, the full Hamiltonian is approximated with a low energy one that is linear around $\boldsymbol{K}$. 
 Here, we show that as long as non-adiabacity is under consideration, quadratic terms could become important corresponding to a passage through the coalescence point.   To this end, we expand  $s(\boldsymbol{k})$ around $\boldsymbol{K}$  up to the second order (i.e., $\boldsymbol{k}\rightarrow \boldsymbol{K}+ \boldsymbol{q}$). Therefore, 
 \begin{equation}
  \begin{aligned} H=& -\frac{3 h a}{2}\left(\begin{array}{ccc}0 & \alpha (q_{y}-i q_{x}) \\ \alpha^*(q_{y}+i q_{x}) & 0\end{array}\right) \\ &-\frac{3 h a^2}{8}\left(\begin{array}{cc}0 & \alpha(q_{y}+iq_{x})^{2} \\ \alpha^*(q_{y}-iq_{x})^{2} & 0\end{array}\right), \end{aligned}
 \end{equation}
where $\alpha=e^{i\frac{2\pi}{3}}$. Since we are interested in studying TPs for passages through the Dirac points, we set $q_y = 0$, $q_x = \lambda$. By eliminating $\alpha$ through an appropriate gauge transformation, the Hamiltonian reads,
\begin{equation}\label{g2}
H(\lambda)=\gamma\left(\begin{array}{cc}
0 & 2i\lambda + a\lambda^2/2 \\
  -2i\lambda + a\lambda^2/2 & 0
\end{array}\right).
\end{equation}
 The corresponding energy dispersion $E_{\pm}= \pm  \gamma|\lambda|  \sqrt{(a \lambda/2)^2+4}$, where $\gamma=3ah/4$, and the touching point occurs at $\lambda=0$. The above-mentioned Hamiltonian could be transformed to 
 a PW-type Hamiltonian through the global gauge transformation $U = \frac{1}{\sqrt {2}}\left(\begin{array}{cc}
e^{i\frac{\pi}{4}} & e^{-i\frac{\pi}{4}} \\
  e^{i\frac{\pi}{4}} & -e^{-i\frac{\pi}{4}}
\end{array}\right)$. The transformed Hamiltonian ($U^{\dagger} H U$) is
\begin{equation}
 \displaystyle H(\lambda)= -\gamma\left(\begin{array}{cc}
   -\frac{a\lambda^2}{2} & 2\lambda \\
  2\lambda &  \frac{a\lambda^2}{2}
\end{array}\right),
\end{equation}
with,
 \begin{equation}\label{chigr}
  \chi(\lambda) = \left( \frac{4a}{a^2\lambda^2 + 16}\right)^2.
 \end{equation}
Obviously, similar to $\displaystyle \chi_{PW}$, $\chi$ does not become zero at the touching point. 
Noticeably, $\chi(0)$ is only a function of lattice constant $a$, thus could be tuned by the strain.
 We should note that, the presence of the phase difference between the linear and quadratic term is essential for $\chi$ to become non-zero at $\lambda=0$.\\
\indent If we keep only the linear term in Eq.~(\ref{g2}), trivially $\chi$ becomes zero. 
 This means that a transition to the excited state occurs with a probability equal to one. 
 However, Eq.~(\ref{chigr}) tells us that the TP to the excited state  becomes 
  less than one, as $\chi(0)\neq 0$ (even in the arbitrary vicinity of the touching point).\\
  \indent It is worth mentioning that when $q_y\neq 0$, by only considering
  the linear terms for $q_x$, the problem will be reduced to a basic LZ problem, where LZ formula gives a fairly good estimation of the tunneling probability \cite{lim2012}.\\
\indent To figure out the effect of larger order terms in the above-mentioned examples we consider the following Hamiltonian,
\begin{eqnarray}\label{eq2}
   H = \left(
\begin{array}{cc}
 g(\lambda)  & \Delta  \lambda  \\
 \Delta  \lambda  & -g(\lambda)  \\
\end{array}
\right)
  \end{eqnarray}
  where $\displaystyle g(\lambda)=\sum_{n=1..N} a_n\lambda^n$, and $\lambda$ and $\Delta$ are real numbers. The corresponding $\chi(\lambda)$ has the following form:
 \begin{eqnarray}\label{chigeneral}
\displaystyle\chi {(\lambda ) = }\left ( \frac{\Delta}{2}\times\frac{ a_2+\lambda k(\lambda)}{ \Delta ^2+\left ({g(\lambda)}/{\lambda}\right)^2}\right )^2,
 \end{eqnarray}
with $\displaystyle k(\lambda)=\sum_{n=2..N-1}na_{n+1}\lambda^{n-2}$.
One can easily check that as $\lambda\rightarrow 0$, then $\chi(\lambda)\propto a_2^2$, which is nonzero as long as $a_2\neq 0$. This means that $\chi(0)$ is only affected by second-order terms. The terms with an exponent larger than two will affect  $\chi$ only away from the touching. 
 \indent To better understand the role of terms with exponents larger than two in Eq.~(\ref{eq2}), one may define
\begin{eqnarray}\label{hamsgl}
{H}_{GL}^{(n)}(\lambda) &=& \displaystyle\left( {\begin{array}{cc} \lambda^n &\lambda\Delta_1 \\ \lambda\Delta_1  &  -\lambda^n\\ \end{array} } \right),
\end{eqnarray}
with,
\begin{eqnarray}
  \displaystyle \chi_{GL}^{(n)}(\lambda)&=&\frac{(n-1)^2 \Delta_1^2 \lambda^{2(n-2)}}{4(\Delta_1^2+\lambda^{2(n-1)})^2}.
\end{eqnarray}\\
In the following section, we will first qualitatively analyze  ${\cal P}^{(n)}_{GL}$ for $n = 1, 2$ where $n=2$ corresponds to $H_{PW}$. For ${\cal P}^{(n>2)}_{GL}$  we observe, in general, two distinct behaviors for even and odd values of $n$. Hereafter, we refer to $\Delta_1$ as the linear coupling.
 % ----------------------------------------------
 
 %=========================================================================
 %%%%%%%%%%%%%%%%%%%%%%%%%%%%%%%%%%%%%%%%%%%%%%%%%%%%%%%

 \section{Qualitative Analyses}  \label{analysis}
In this section, we qualitatively analyze the behavior of TPs for each
 ${ H_{GL}^{(n)}(\lambda)}$ for different values of $n$.
 As we will see in our models, the dispersion of $\chi(\lambda)$ consists of different peaks. Therefore, for the sake of simplicity, we label the location of the peaks with MFP (maximum fidelity point).
 For the location of the minimum gap, we use MGP (minimum gap point). \\
 %%% 
 \\
\indent  ${\bf H_{GL}^{(1)}(\lambda)}$: This model is gap-less with $\lambda_{MGP}=0$, and $\chi(\lambda)=0$ everywhere.
  Therefore, by passing through the crossing point, the transition to the excited state becomes one for both adiabatic and nonadiabatic regimes, with no change in the character of the initial state. \\

 %=========================================================================
  %=========================================================================

\begin{figure}[h]%
 \includegraphics*[width=0.45\textwidth]{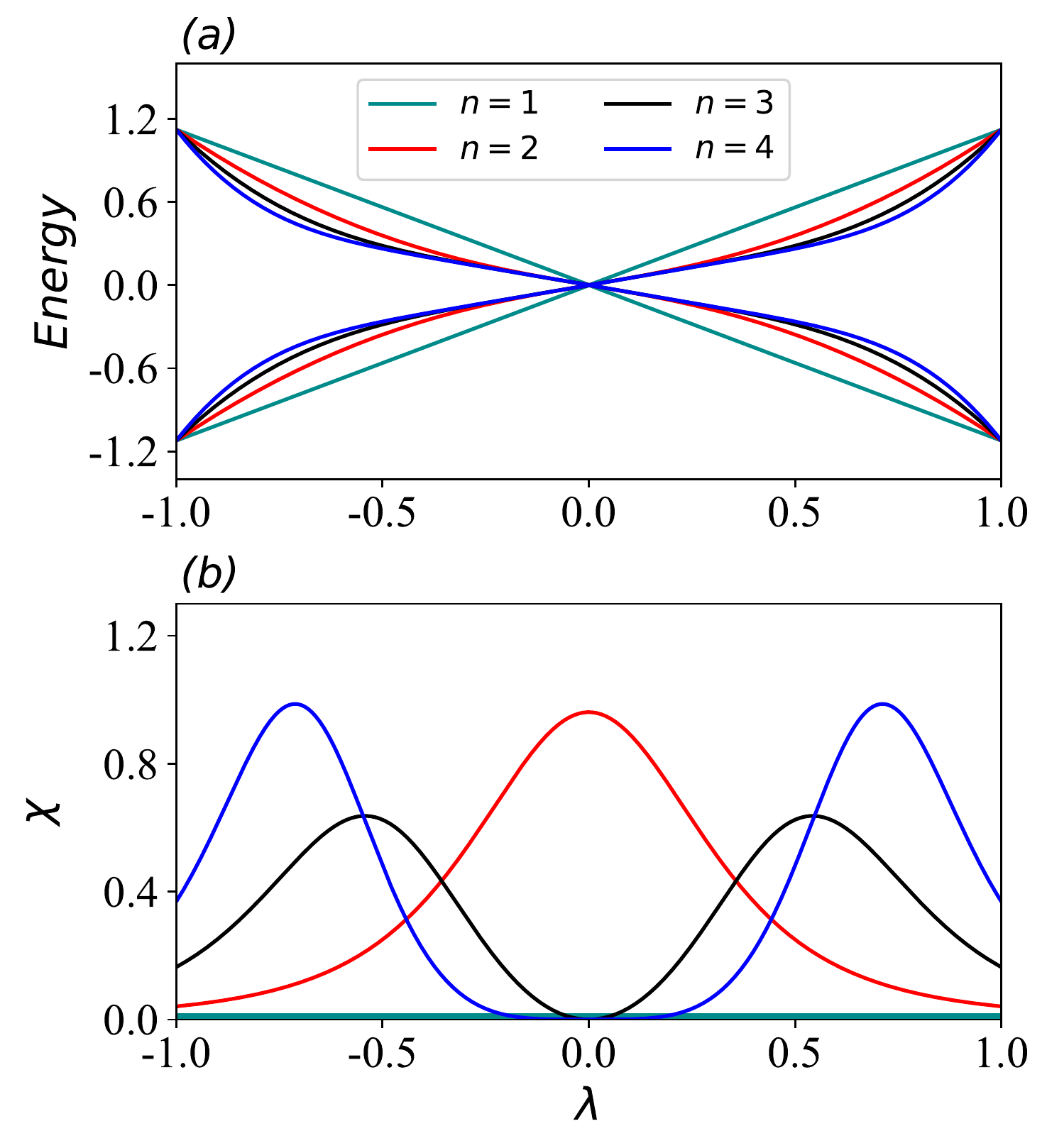}%
  \caption[]{%
     (Color line) (a) Energy dispersion as a function of parameter $\lambda$. (b) The FIS as a function of $\lambda$ for different orders of diagonal part of gap-less Hamiltonian, $H^{(n)}_{GL}$ and  for $\Delta_1=0.5$.}
    \label{fig1}
\end{figure}

  \indent ${\bf H_{GL}^{(2)}(\lambda)}$ ({\bf 1D p-wave like Hamiltonian}): Here, $\lambda_{MGP}=\lambda_{MFP}=0$ (see Fig.\ref{fig1}), and
   $\chi(\lambda)=\frac{\Delta_1^2}{4(\lambda^2+\Delta_1^2)^2}$ (which is similar to $\chi(\lambda)$ of the LZ).  
   Since $\chi(0)\neq 0$, therefore different from $H_{GL}^{(1)}$, the nonadiabatic transition is not equal to one for a finite speed of the driving of the system through the touching point. Here, the faster one drives the system, the probability that the  evolved state remains in the instantaneous ground-state becomes larger, and there is no change of the character for fast enough speeds (for numerical results see Sec.~\ref{numtest}). \\
   \indent  For an adiabatic evolution, the evolved state acquires a change of the character by passing through the crossing point, and a transition to the excited state occurs with a TP equal to one. As explained in Sec.~\ref{models}, $H_{GL}^{(2)}$ is not reducible to a linear Hamiltonian in terms of both adiabatic and nonadiabatic evolution.\\ 

            \begin{figure}[h]%
 \includegraphics*[width=0.45\textwidth]{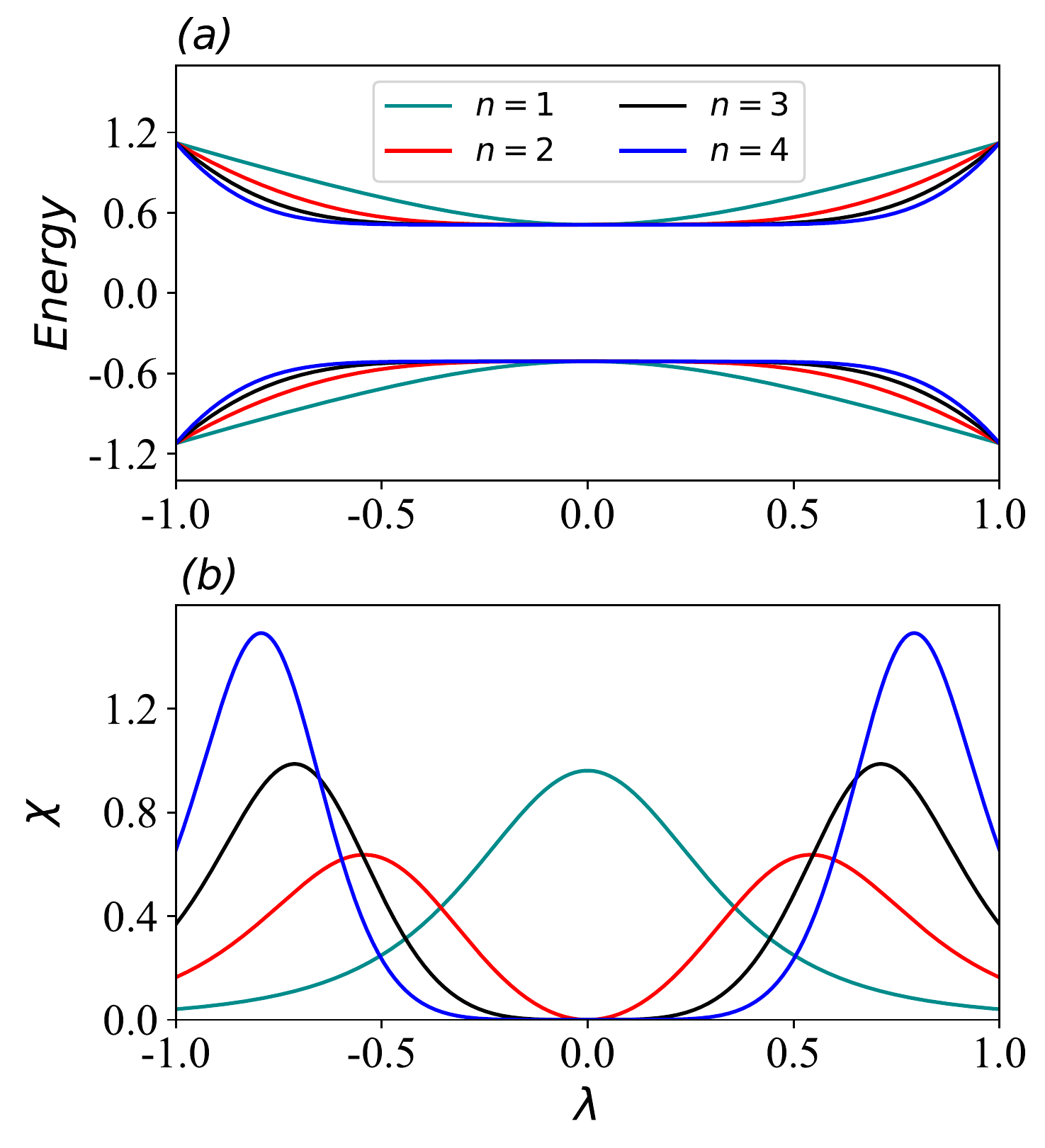}%
  \caption[]{%
     (Color line) (a) Energy dispersion as function of parameter $\lambda$, (b) The FIS as function of $\lambda$ for different orders of diagonal part of gapped Hamiltonian, $H^{(n)}_{GP}$ and  for $\Delta_2=0.5$.}
    \label{fig2}
\end{figure}
 %
 %=========================================================================

\indent ${\bf H_{GL}^{(n=odd)}(\lambda)}$: In this case $\chi$ acquires a two peak structure 
 as a function of $\lambda$ with $\lambda_{MGP}=0$ and $\lambda_{\pm, MFP}=\pm\left( \frac{n-2}{n}\Delta_1^2\right)^{\frac{1}{2(n-1)}}$ (see Fig.\ref{fig1}).
%Trivially,$\lambda_{MFP}\ne \lambda_{MGP}$. 
 Different from $H_{GL}^{(2)}$, $\chi(\lambda)$ acquires its minimum at $\lambda_{MGP}$, i.e., $\chi(\lambda_{MGP}) = 0$. 
 That is, upon passing through the touching point with a PL$_1$ protocol, the transition to the excited state almost happens without a change of the character of the evolved wave function. In other words, as long as the starting state of the evolution process is the ground-state of the system which is close enough to the touching point, the system is reducible to a simple crossing Hamiltonian in terms of nonadiabatic transition (i.e., ignoring the diagonal cubic terms in the Hamiltonian might be a good approximation).\\
 \indent For a PL$_2$ protocol, apart from the physical relevance of such a model, because $H_{GL}^{(n=odd)}(-\lambda) = -H_{GL}^{(n=odd)}(\lambda)$, the wavefunction propagates back in time by passing through the touching point. Therefore, a transition to the excited state with a probability equal to one is warranted (without a change of the character). %
 % at $t=+\infty$ for arbitrary couplings and speed of driving.\\

 %=========================================================================
 %=========================================================================

 \indent ${\bf H_{GL}^{(n=even)}(\lambda)}$: This situation is very similar to what was discussed about $H^{(n=odd)}_{GL}$. That is,  $H^{(n=even)}_{GL}$ is reducible to a simple crossing model close enough to the touching point. However, different from $H_{GL}^{(n=odd)}$, the $H_{GL}^{(n=even)}(-\lambda)\neq -H_{GL}^{(n=even)}(\lambda)$. Therefore, the 
   evolution of the system is not as trivial as odd cases. \\ 
  \indent If the starting point of the evolution process is considered far away from the touching point, i.e., beyond one of the MFP points, then upon nonadiabatic driving of the system through the MGP (a PL$_2$ protocol), the  TP can be understood at two asymptotic limits of large and small values of 
   $\Delta_1$.\\
  \indent When $\Delta_1 \ll 1$ we have $\displaystyle c\sqrt{\chi} \gg1$.  Then, the system is highly nonadiabatic and the evolved wave function preserves its initial form which results in ${\cal P} \ll 1$.\\
  \indent On the other hand, $\displaystyle c\sqrt{\chi} \ll 1$ when $\Delta_1 \gg 1$ (adiabatic limit). In this limit, the evolved wave function will have a large contribution from the ground-state (the excited state) for $\lambda(t)<0$ ($\lambda(t)>0$). Consequently, ${\cal P}\sim 1$, and a change of character from $ |\psi(-\infty)\rangle=\left( \begin{array}{cc} 0 \\ 1\end{array}\right)$ to $|\psi(+\infty)\rangle=\left( \begin{array}{cc} 1 \\ 0\end{array}\right)$ occurs.
  \indent For the intermediate values of $\Delta_1$, ${\cal P}$ interpolates between these two limits.\\
 %=========================================================================
 %=========================================================================
 %
  \begin{figure}[h]%
 \includegraphics*[width=0.45\textwidth]{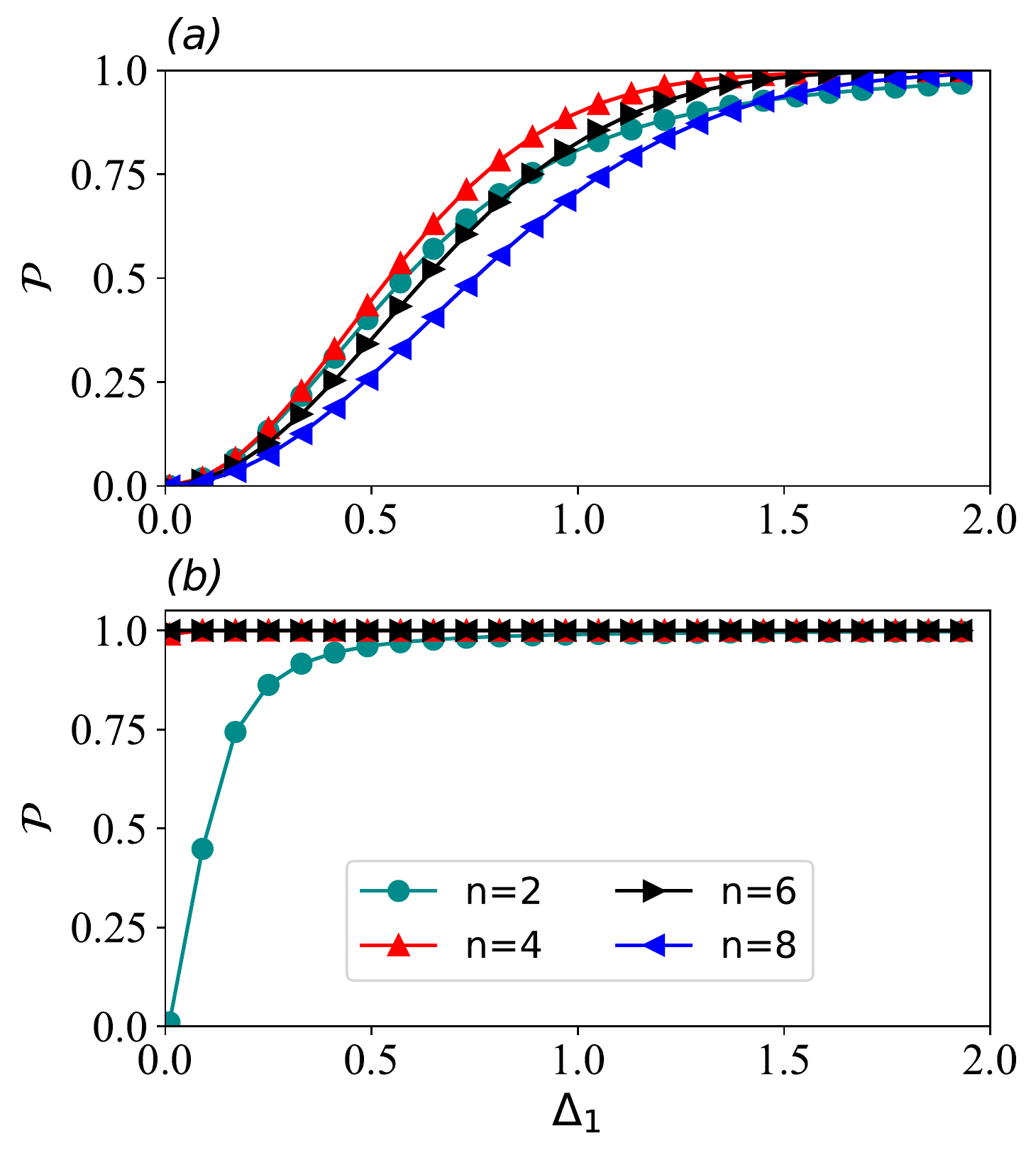}%
  \caption[]{%
     (Color line) The ${\cal P}^{(n)}_{GL}$ as a function of the linear coupling $\Delta_1$, with $n=2,4,6,8$ and $c=0.1$, for PL$_2$ (a) and  PL$_1$ (b) protocols.} 
    \label{fig_tp_gl_infty_01}
\end{figure}  
% ######################
% #####################
% we say sth about gapped

%%%%%%%%%%%%%%%%%%%%%%%%%%%%%%%%%%%%%%%%%%%%%%%%%%%%%%%
%=========================================================================
%%%%%%%%%%%%%%%%%%%%%%%%%%%%%%%%%%%%%%%%%%%%%%%%%%%%%%%

\section{Numerical results}\label{numtest}
In order to complement the analyses in the previous section, here we provide our numerical findings for TP.  As explained in Sec.~\ref{analysis}, $H^{(n=odd)}_{GL}$ has a trivial time evolution behavior. Hence, we ignore it in the remainder of our discussion.\\
\indent Since, in the context of nonadiabatic transitions, gapped systems have been well studied in the past, it would be a good practice to compare our results for gap-less models with those of gapped systems with the same order of diagonal terms. To this end, we define \cite{suominen1992parabolic,PhysRevA.88.043404,PhysRevA.59.4580},
 \begin{eqnarray}\label{hamsgp}
{H}_{GP}^{(n)}(\lambda) &=& \displaystyle\left( {\begin{array}{cc} \lambda^n &\Delta_2\\ \Delta_2 &-\lambda^n   \\ \end{array} } \right),
\end{eqnarray}
for which,
 \begin{eqnarray}
  \displaystyle \chi_{GP}^{(n)}(\lambda)&=&\frac{n^2\Delta_2^2 \lambda^{2(n-1)}}{4(\Delta_2^2+ \lambda^{2n})^2},
 \end{eqnarray}
 with $\lambda_{MGP}=0$ and $\lambda_{\pm,MFP}=\pm\left( \frac{n-1}{n+1}\Delta_1^2\right)^{\frac{1}{2n}}$ (see Fig.\ref{fig2}). Setting $n=1$, $H^{(1)}_{GP}$ is nothing but a basic Landau-Zener Hamiltonian. One can also check that $ \chi_{GP}^{(n-1)}\equiv \chi_{GL}^{(n)}$.\\
\indent Similar to the Hamiltonian of each model, the TP is labeled with the GL/GP and the order $n$. The details of the method employed for the time evolution, the definition of TP (${\cal P}$) and the definition of driving protocols are explained in Sec.~\ref{method}.  
\begin{figure}[h]%
 \includegraphics*[width=0.45\textwidth]{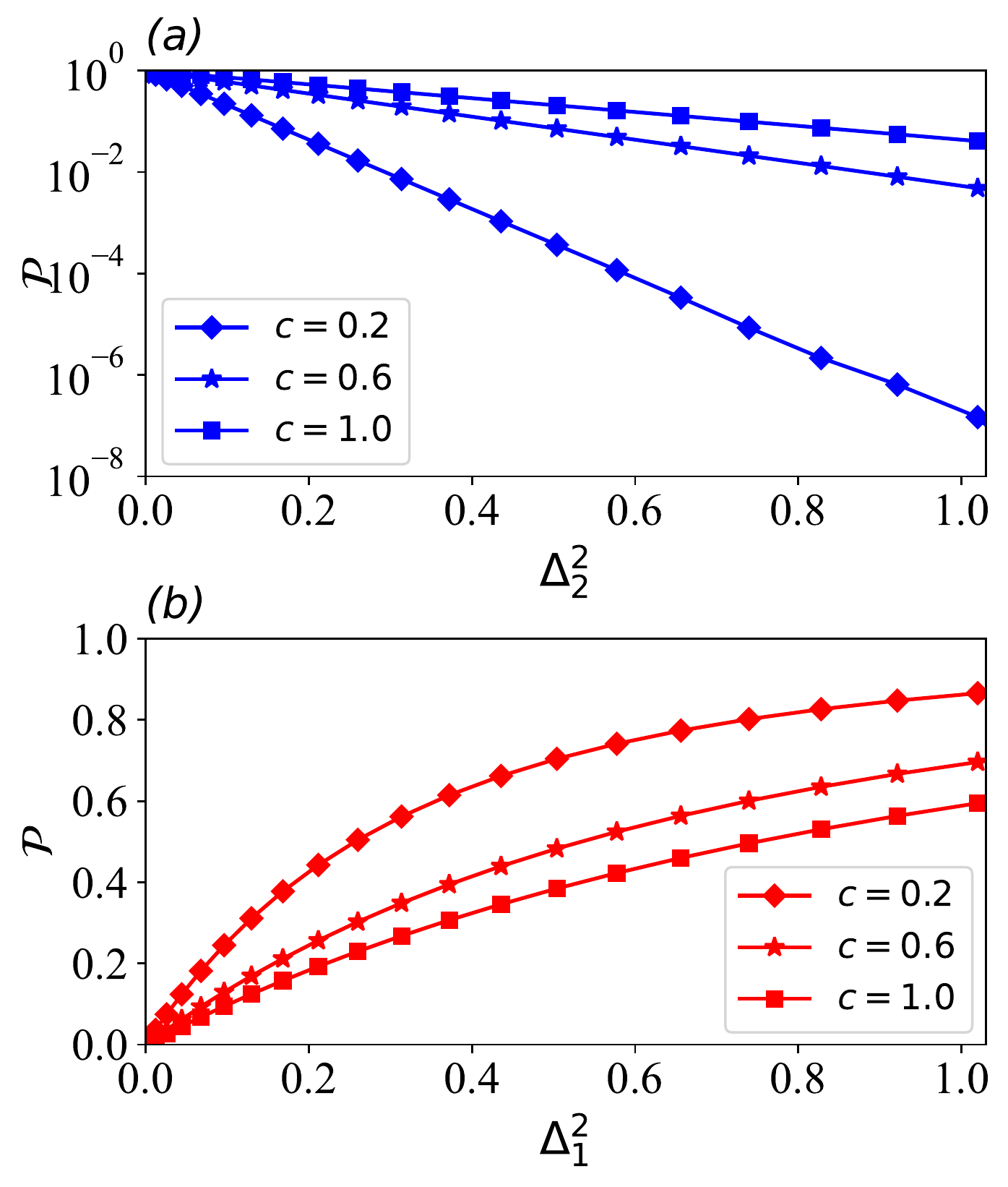}%
  \caption[]{%
     (Color line) (a) ${\cal P}^{(1)}_{GP}$ as a function of $\Delta_2^2$ in log-scale and for different speeds of driving $c$. (b) ${\cal P}^{(2)}_{GL}$  as a function of $\Delta_1^2$ for different speeds of driving $c$. Solid lines are guides for eye.
     }
    \label{fig4}
\end{figure}    
\indent In Fig.~\ref{fig_tp_gl_infty_01}(a), ${\cal P}^{(n)}_{GL}$ is plotted as a function of $\Delta_1$, with a PL$_2$ protocol and for different values of $n=2,4,6,8$. For all cases, by increasing $\Delta_1$, ${\cal P}^{(n)}_{GL}$ monotonously is enhanced, which is in accord with our qualitative analyses in section.~\ref{analysis}. That is, by increasing $\Delta_1$, FIS is suppressed. Consequently, the system approaches the adiabatic 
 regime which results in a TP equall to one.  \\
 \indent In Fig.~\ref{fig_tp_gl_infty_01}(b), we have the same plot as Fig.~\ref{fig_tp_gl_infty_01}(a) but for a 
 PL$_1$ protocol. Obviously, ${\cal P}^{(n)}_{GL}\simeq 1$ 
  for $n>2$ cases, which means non-linear diagonal terms ($\lambda^n$) have a very marginal effect on the TP as long as the starting point is close enough to the touching point, which is the case for a PL$_1$ protocol. This is also in agreement with our qualitative analysis. That is, near the crossing point, the FIS becomes very small and a complete TP to the excited state is expected. However, ${\cal P}^{(2)}_{GL}$ shows a different behavior for a PL$_1$ protocol in comparison with $n>2$ cases. In this case when $\Delta_1\rightarrow 0$
     , ${\cal P}^{(2)}_{GL}\rightarrow 0$. The reason for such behavior is that when $\lambda\rightarrow 0 $ then $\chi\rightarrow \frac{1}{\Delta_1}$. Consequently, $\chi$ is enhanced when $\Delta_1\rightarrow 0$. This means the system is still highly nonadiabatic for small values of $\Delta_1$ which finally results in a suppressed transition to excited state for small values of $\Delta_1$.\\
\indent In Fig.\ref{fig4}(a), ${\cal P}^{(1)}_{GP}$ ($\equiv{\cal P}_{LZ}$) is plotted versus ${\Delta^2_2}$ and for different values of $c$. As expected, it decays exponentially as a function of ${\Delta^2_2}$ and decreases as $c$ becomes smaller. \\
\indent Fig.\ref{fig4}(b) shows ${\cal P}^{(2)}_{GL}$ versus $\Delta^2_1$ and for different values of $c$. The figure depicts that ${\cal P}^{(2)}_{GL}$ is increased simultaneous to the enhancement of $\Delta_1^2$, and is suppressed upon the amplification of $c$ which is in stark difference with ${\cal P}^{(1)}_{GP}$. Interestingly, ${\cal P}^{(2)}_{GL}$ 
  reveals that the evolved wave function tends to have larger ground-state contribution by increasing the speed of driving.\\
%
 %%%%%%%%%%%%%%%%%%%%%%%%%%%%%%%%%%%%%%%%%%%%%%%%%%%%%%%
 %%%%%%%%%%%%%%%%%%%%%%###########################
  \begin{figure}[h]%
 \includegraphics*[width=0.45\textwidth]{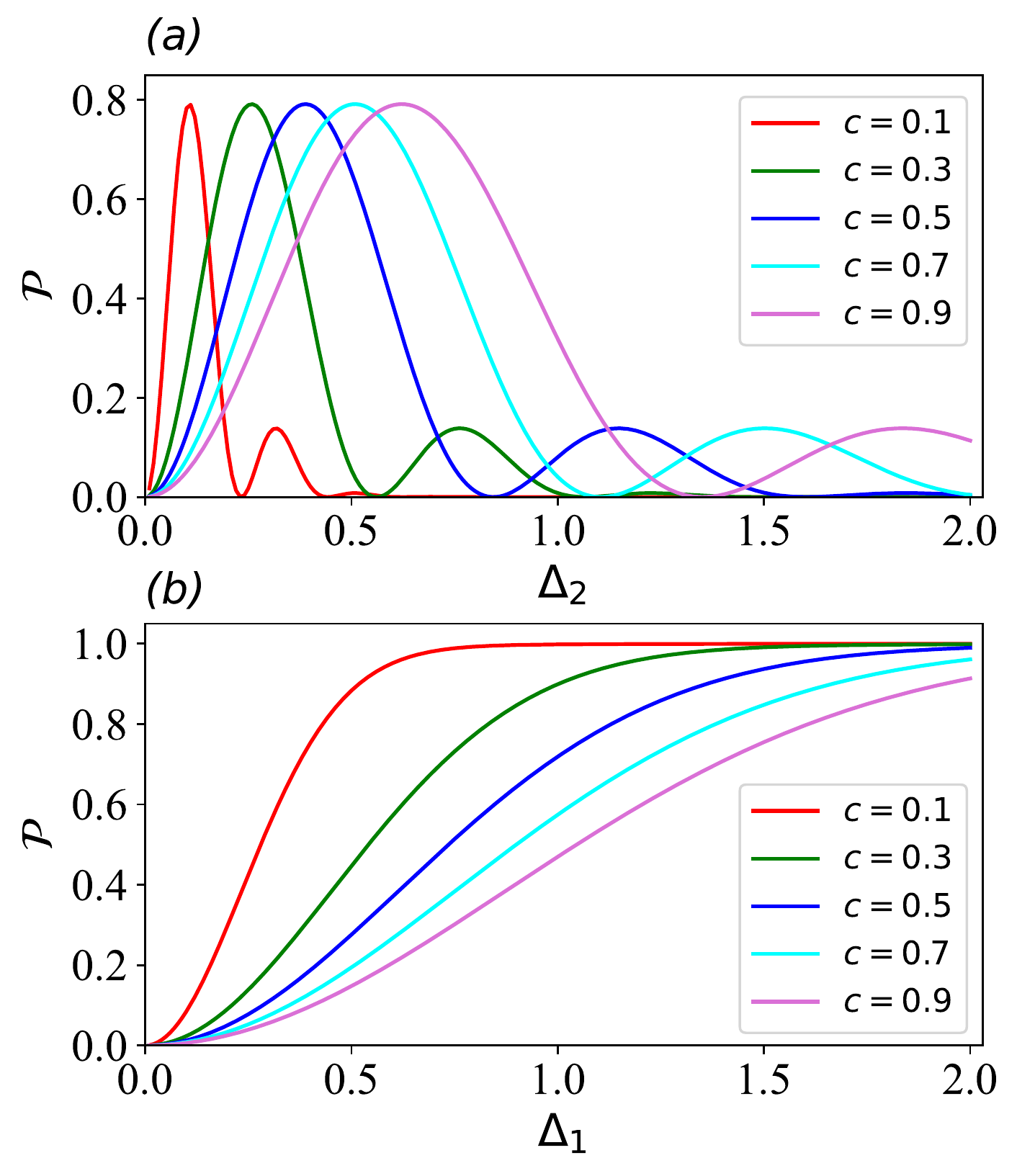}
  \caption[]{%
     (Color line) (a) Transition probability, ${\cal P}^{(4)}_{GP}$, as a function of $\Delta_2$ and different speeds of driving $c$ for a PL$_2$ protocol. (b) ${\cal P}^{(4)}_{GL}$, as a function of $\Delta_1$ and different speeds of driving $c$ for a  PL$_2$ protocol.} 
    \label{fig_tp_gp4_gl4}
\end{figure}
 %%%%%%%%%%%%%%%%%%%%%%%%%%%%%%%%%%%%%%%%%%%%%%%%%%%%%%%
%%%%%%%%%%%%%%%%%%%%%%%%%%%%%%%%%%%%%%%%%%%%%%%%%%%%%%%%%%%
 \begin{figure}[h]%
 \includegraphics*[width=0.45\textwidth]{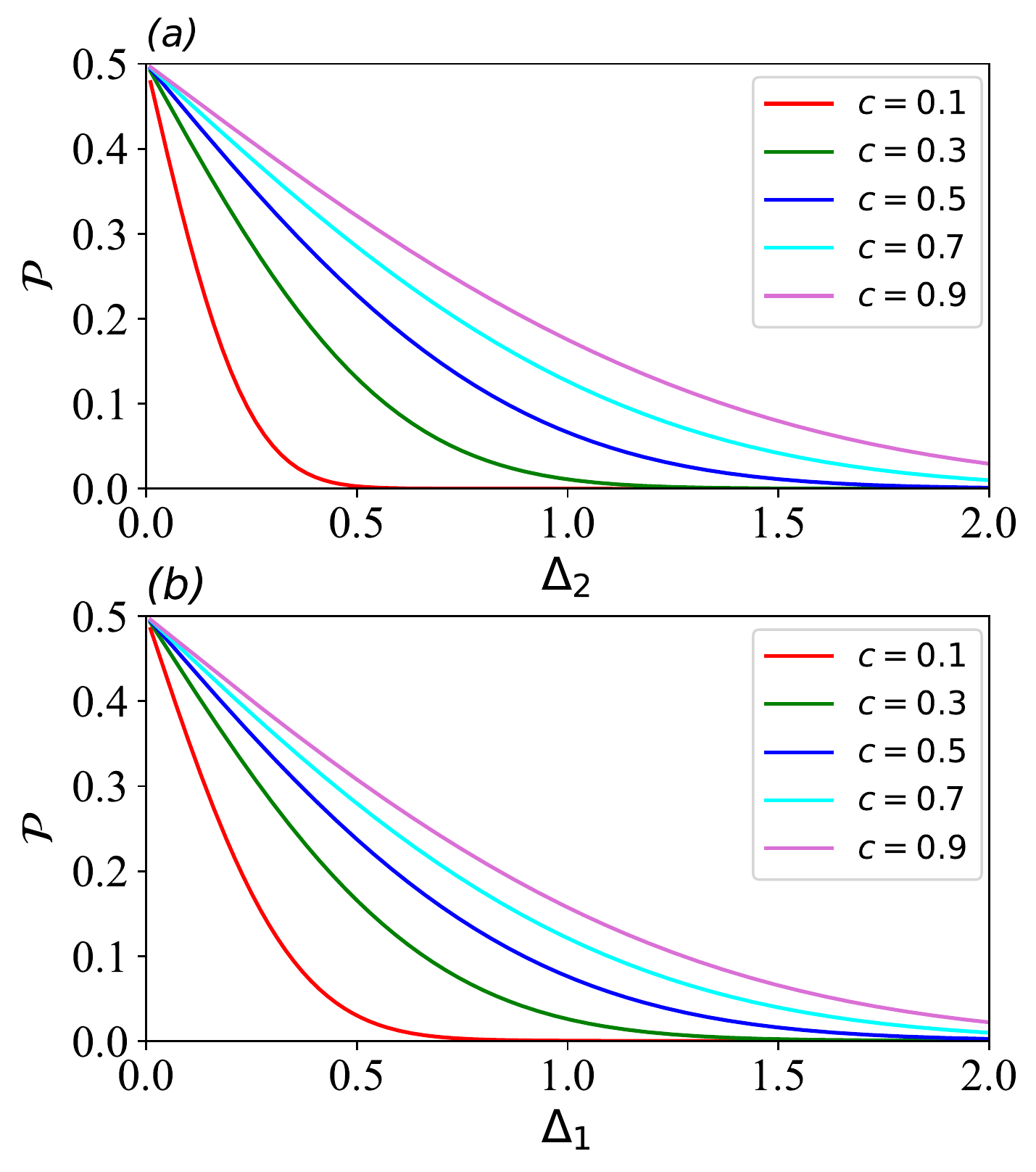}%
  \caption[]{%
     (Color line) (a)  ${\cal P}^{(4)}_{GP}$ as a function of $\Delta_2$, for a PL$_{-\infty,0}$ protocol with different speeds of driving $c$. (b) ${\cal P}^{(4)}_{GL}$ as a function of the linear coupling $\Delta_1$ for PL$_{-\infty,0}$ protocol with different speeds of driving $c$.} 
    \label{fig_tp_glgp_to0}
\end{figure} 
 %%%%%%%%%%%%%%%%%%%%%%%%%%%%%%%%%%%%%%%%%%%%%%%%%%%%%%%
 \indent As discussed before, the TP for gap-less models shows a monotonous enhancement towards one as a function of the linear coupling.\\
 \indent Now we compare the TP of gapped and gap-less models with the same order  $n$. In Fig.\ref{fig_tp_gp4_gl4}(a), ${\cal P}^{(4)}_{GP}$ is plotted as a function of $\Delta_2$, as it is expected from semiclassical analysis of the TP\cite{suominen1992parabolic,PhysRevA.86.033415}, an oscillating behavior as a function of $\Delta_2$ is observed. Based on the DPP method, such oscillation arises from the zeros of the energy difference  between the two states. Such zeros consist of both real and imaginary parts which finally result in an oscillatory behavior in ${\cal P}^{(4)}_{GP}$.\cite{PhysRevA.86.033415} The oscillatory behavior prevails for ${\cal P}^{(n)}_{GP}$ with arbitrary $n>1$. 
  \indent In Fig.\ref{fig_tp_gp4_gl4}(b), we also depict ${\cal P}^{(4)}_{GL}$ as a function of the linear coupling and for different speeds of driving, which shows a  monotonous enhancement of ${\cal P}^{(4)}_{GL}$ to one as a function of $\Delta_1$ and $1/c$. \\
  \indent Oscillations of ${\cal P}^{(4)}_{GP}$ as a function of $\Delta_2$ could be explained based on a real-time analysis, through a gap dependent inteference of the two peaks in the dispersion of $\chi$ for a PL$_2$ protocol, where a passage through both peaks occurs. Moreover, monotonous ${\cal P}^{(4)}_{GL}$ as a function of $\Delta_1$ behavior is a signature of a fixed interference between the two peaks, which does not depends on the $\Delta_1$. This is generally correct for ${\cal P}^{(n)}_{GL}$ with even $n>2$.\\ 
  
 %%%%%%%%%%%%%%%%%%%%%%%%%%%%%%%%%%%%%%%%%%%%%%%%%%%%%%%
 \indent To elaborate on the inteference of the two peaks in gapped (for $n>1$) and gapless models (for $n>2$), since there are two major peaks at $\lambda_{\pm,MFP}$ separated by a region with $\chi\simeq 0$ in the spectrum of $\chi$ (as a function of $\lambda$), we divide PL$_2$, into a PL$_{-\infty,0}$ followed by a PL$_{0,\infty}$. We first compare $\displaystyle {\cal P}^{(4)}_{GL}$ and $\displaystyle {\cal P}^{(4)}_{GP}$ for a PL$_{-\infty,0}$ protocol, which consists of passing through $\lambda_{-,MFP}$. In practice, for the case of gap-less models we evolve from $-\infty$ to $-\epsilon$, where $\epsilon$ is an infinitesimal positive number. In Fig.~\ref{fig_tp_glgp_to0}(a), for a PL$_{-\infty,0}$ driving, the ${\cal P}^{(4)}_{GP}$ is depicted as a function of $\Delta_2$  and for different speeds of driving. A clear observation is that the TP monotonously fades as a function of $\Delta_2$. The same is observed for ${\cal P}^{(4)}_{GL}$ subjected to a PL$_{-\infty,0}$ driving as a function of the linear coupling $\Delta_1$. The latter is depicted in Fig.~\ref{fig_tp_glgp_to0} (b). Clearly, $\cal P$ behaves very similarly for both GP and GL cases and there is no oscillations in the ${\cal P}^{(4)}_{GP}$ and ${\cal P}^{(4)}_{GL}$ as a function of $\Delta_2$ and $\Delta_1$, respectively.\\
%%%%%%%%%%%%%%%%%%%%%%%%%%%%%%%%%%%%%%%%%%%%%%%%%%%%%%%
      \indent Next we look at the effect of PL$_{0,\infty}$ on the result of PL$_{-\infty,0}$. This could be explained as follows. Starting from the instantaneous ground-state at $t= -\infty$, we evolve the wave function from $t = -\infty$ to $t = 0$, where $\chi\simeq 0$ in this region. Then, by writing the evolved wave function based on instantaneous eigenstates at $t=0$, we evolve each of the instantaneous eigenstates to $t=+\infty$, the final evolved state at $t=+\infty$ is the superposition of the evolved instantaneous eigenstates at $t=0$. The result of the TP must be exactly the same as the single PL$_2$ protocol. This could be written as follows:
     \onecolumngrid
       \begin{eqnarray}
        |\psi(t=+\infty)\rangle &=& U(+\infty,0)U(0,-\infty)|\phi_0(t=-\infty)\rangle 
                               = U(+\infty,0)(\alpha_+|\phi_1(0)\rangle +  \alpha_-|\phi_0(0)\rangle   ) \nonumber\\ 
                               &=& (\alpha_+\beta_+^+ + \alpha_-\beta_-^+)|\phi_1(+\infty)\rangle +  (\alpha_+\beta_+^- + \alpha_-\beta_-^-) |\phi_0(+\infty)\rangle,
       \end{eqnarray}
   \twocolumngrid
where $\alpha_{\pm}=\langle \phi_{1,0}(0) |U(0,-\infty)|\phi_0(-\infty)\rangle$,
 $\beta_{+}^{\pm} = \langle \phi_{1,0}(+\infty) |U(+\infty,0)|\phi_1(0)\rangle$ and 
  $\beta_{-}^{\pm} = \langle \phi_{1,0}(+\infty) |U(+\infty,0)|\phi_0(0)\rangle$. Therefore TP ${\cal P}=|\alpha_+\beta_+^+ + \alpha_-\beta_-^+|^2$. By writing 
    $\alpha_\pm= |\alpha_\pm|e^{i a_{\pm}}$, $\beta_{\pm}^{+}=|\beta_{\pm}^+|e^{i b_{\pm}}$, and 
    $ \beta_{\pm}^{-}=|\beta_{\pm}^{-}|e^{i \tilde{b}_{\pm}}$, the TP could be written as
    \begin{eqnarray}
     {\cal P}_{tr} &=& |\alpha_+\beta_+^+|^2 + |\alpha_-\beta_-^+|^2\\ &+&
      2 |\alpha_+\beta_+^+|\times|\alpha_-\beta_-^+| cos \Delta\Phi
    \end{eqnarray}
    with $\Delta\Phi= \Im \ln [\alpha_+\beta_+^+(\alpha_-\beta_-^+)^*] = [a_++b_+-a_--b_-]_{mod~2\pi}$, and $\Xi_{mod~2\pi}=\Xi~mod~2\pi ~-\pi$. The $\Delta\Phi$ is the indicator of the interference of the two peaks in the spectrum of $\chi$.\\
%%%%%%%%%%%%%%%%%%%%%%%%%%%%%%%%%%%%%%%%%%%%%%%%%%
     \begin{figure}[h]%
 \includegraphics*[width=0.5\textwidth]{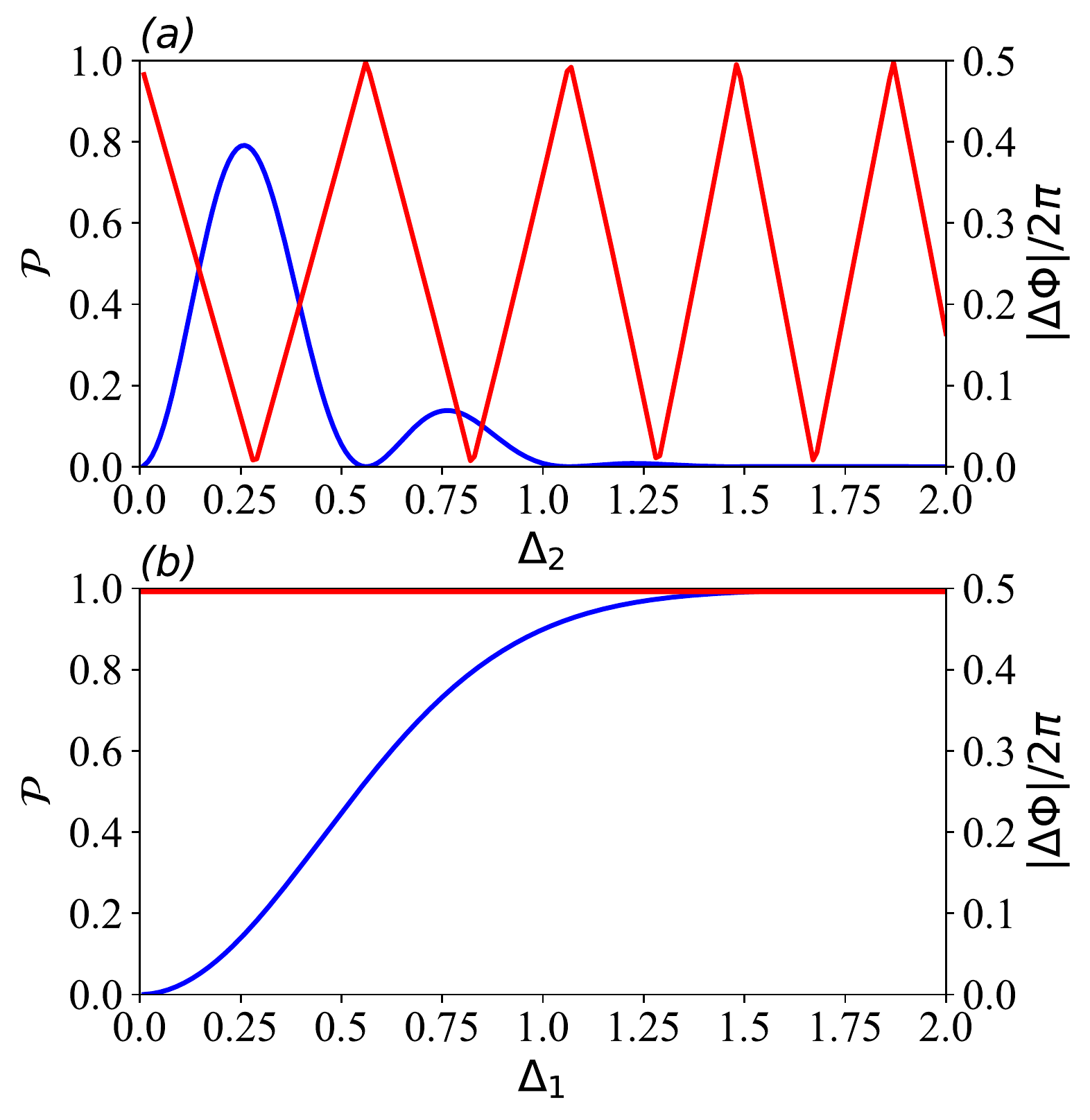}%
  \caption[]{%
     (Color line) (a)  ${\cal P}^{(4)}_{GP}$ (blue) and $|\Delta\Phi|/2\pi$ (red)  as a function of coupling $\Delta_2$, for a PL$_2$ protocol with $c=0.3$. (b) ${\cal P}^{(4)}_{GL}$ (blue) and $|\Delta\Phi|/2\pi$ (red) as a function of the linear coupling $\Delta_1$, for a PL$_2$ protocol with $c=0.3$.} 
    \label{fig_tp_phase_glgp4}
\end{figure} 
%%%%%%%%%%%%%%%%%%%%%%%%%%%%%%%%%%%%%%%%%%%%%%%%%%
 \begin{figure}[h]%
 \includegraphics*[width=0.42\textwidth]{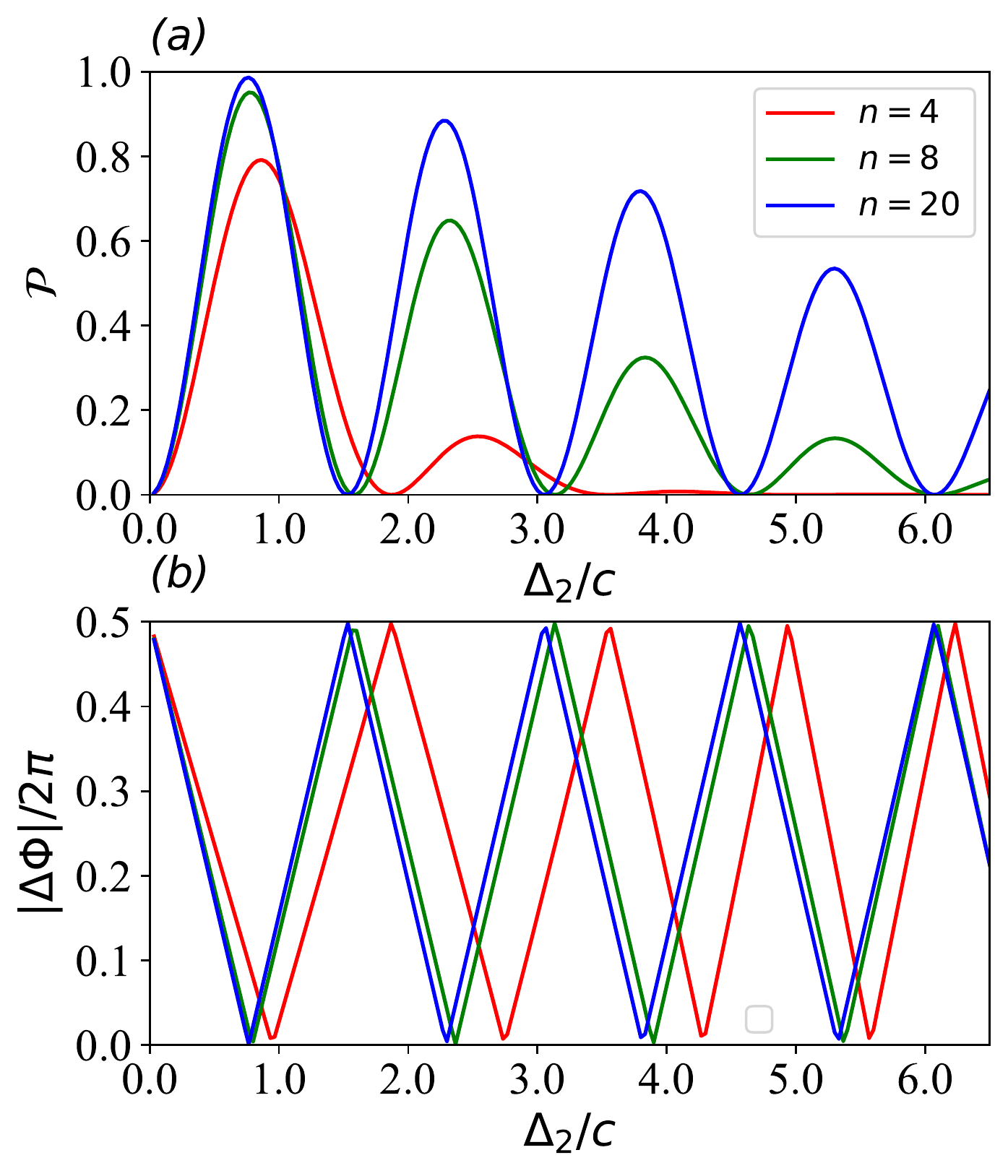}%
  \caption[]{%
     (Color line) (a)  ${\cal P}^{(n)}_{GP}$ as a function of   $\Delta_2/c$, for a PL$_2$ protocol with  $n=4,8,20$. (b) The corresponding $|\Delta\Phi|/2\pi$ as a function of $\Delta_2/c$.} 
    \label{fig_tp_gp4_8_20_phase}
\end{figure}  
%%%%%%%%%%%%%%%%%%%%%%%%%%%%%%%%%%%%%%%%%%%%%%%%%%
 \indent In Fig.~\ref{fig_tp_phase_glgp4}(a), ${\cal P}^{(4)}_{GP}$ is plotted for a PL$_2$=PL$_{-\infty,0}$+PL$_{0,\infty}$ protocol. As it is expected, we observe an oscillatory behavior of the TP as a function of $\Delta_2$. Moreover $|\Delta\Phi|/2\pi$ is also plotted as a function of $\Delta_2$ which shows a linear dependence as a function of $\Delta_2$. In the range that $\Delta\Phi$ oscillates, i.e., between $-\pi$ and $\pi$, ${\cal P}$ attains its maximum  at places where $\Delta\Phi\simeq 0$ and becomes zero at $|\Delta\Phi|=\pi$. \\
     \indent  In Fig.~\ref{fig_tp_phase_glgp4}(b), ${\cal P}^{(4)}_{GL}$ is plotted as a function of the linear coupling $\Delta_1$. In stark difference with gapped models, ${\cal P}^{(4)}_{GL}$ does not show any change of interference between the two peaks since $|\Delta\Phi|=\pi$ is constant as a function of $\Delta_1$. This leads to the absence of  oscillations for the ${\cal P}$ as a function of $\Delta_1$. \\
 \indent The main reason for the linear dependence of $\Delta\Phi$ as a function of $\Delta_2$ for gapped models could be understood as follows. For a PL$_2$ protocol, there is a region between $\lambda_{\pm,MFP}$, where $\chi\simeq 0$ and there is no transition. If one writes the evolved state based on eigenstates in this region, the only thing that happens in this region is the accumulation of phases for each instantaneous eigenstates during the evolution. Since, the dispersion of energies is almost flat in this region (see Fig.~\ref{fig2}), the phase difference between two states becomes $\Delta\Phi=[\int_{-\lambda_{-,MFP}}^{\lambda_{+,MFP}} \frac{(E_2-E_1)}{c}d\lambda]_{mod~2\pi}=[4\lambda_{+,MFP}\Delta_2/c]_{mod~2\pi}$. When $n>>1$ then $\lambda_{+,MFP}\rightarrow 1$. Therefore $\Delta\Phi= [4\Delta_2/c ]_{mod~ 2\pi}$, is expected. For smaller values of $n$ the region with $\chi\simeq 0$ shrinks, consequently phase and probability oscillations show a larger period. In Fig.~\ref{fig_tp_gp4_8_20_phase}(a) we plot ${\cal P}^{(n)}$ for different values of $n$ as a function of $\Delta_2/c$. In Fig.~\ref{fig_tp_gp4_8_20_phase}(b) the same is depicted for $|\Delta\Phi|/2\pi$. The period of oscillations approaches to $\pi/2$ for larger $n$, where it is slightly larger for $n=4$ due to the shrunk $\chi\simeq 0$ region in comparison to $n=20$, which is in agreement with our analysis.\\
    \indent The absence of the change of interference in gap-less models could be understood based on the fact that the levels cross and the sign of energy levels changes symmetrically by a passage through  the touching point. Thus dynamical phase accumulations cancel out in the region of small $\chi$, and the interference will not depend on the linear coupling.\\
    
%%%%%%%%%%%%%%%%%%%%%%%%%%%%%%%%%%%%%%%%%%%%%%%%%%
 \section{Conclusion}\label{conclusion}
     In conclusion, by looking at FIS ($\chi$), we investigated the nonadiabatic transition for gapless TLSs with a single touching point in their spectrum as a function of some parameter $\lambda$. We showed that as long as there exists certain types of quadratic terms in the Hamiltonian (as a function of $\lambda$), the low energy form could not be approximated with a linear one even in the arbitrary vicinity of the touching point. That is $\chi(\lambda)\neq 0$ as $\lambda\rightarrow 0$.  This results in ${\cal P}$ (the TP) becoming less than one by passing through the touching point. \\
      \indent We further showed that in the absence of quadratic terms, the low-energy Hamiltonian could be well approximated with a linear one in the sufficient vicinity of the touching point (even with the presence of terms with exponents larger than two). Here, $~\chi(\lambda)\rightarrow 0$ as $\lambda\rightarrow 0$. Noticeably,  terms with larger exponents than two do affect $\chi$ away from the touching point through the extant of a double peak in the dispersion of $\chi$ as a function of $\lambda$ (please see the main text for the definition of the exact form of the Hamiltonians we have considered). In comparison with gapped models, the TP does not show any oscillation as a function of the linear coupling, which is the signature of the absence of the change of interference between two peaks. The reason for the lack of oscillations in this case, is the change of the sign of the energies associated to instantaneous eigenstates by passing through the touching point. Consequently, phase accumulation does not occur, which in turn leads to the absence of the change of interference between two peaks as a function of the linear coupling.
\begin{acknowledgments}
We thank Abolhassan Vaezi for helpful discussions
and carefully reading our paper.
\end{acknowledgments}

 %%%%%%%%%%%%%%%%%%%%%%%%%%%%%%%%%%%%%%%%%%%%%%%%%%%%%%%
 %=========================================================================
 %%%%%%%%%%%%%%%%%%%%%%%%%%%%%%%%%%%%%%%%%%%%%%%%%%%%%%%
\nocite{apsrev41Control}
\bibliographystyle{apsrev4-1}
\bibliography{main.bib}
\end{document}